# Segmentation of pores within concrete-epoxy interface using synchronous chemical composition mapping and backscattered electron imaging

Andrzej M. Żak [a]*, Anna Wieczorek [b], Agnieszka Chowaniec [c], Lukasz Sadowski [c]

[a] Faculty of Mechanical Engineering, Wroclaw University of Science and Technology, Wroclaw, Poland
[b] Faculty of Pure and Applied Mathematics, Wroclaw University of Science and Technology, Wroclaw, Poland
[c] Faculty of Civil Engineering, Wroclaw University of Science and Technology, Wroclaw, Poland
* Corresponding author: andrzej.zak@pwr.edu.pl

**Abstract**

The method of assessing porosity using images from scanning electron microscopy is ineffective in situations where the substrate and the coating have a significantly different average atomic number, which results in a different contrast of backscattered electrons. For this reason, the analyzes carried out to date were usually based on the manual distinction of important microstructure elements, which was followed by selective thresholding. However, this method depends on the subjective action of the researcher, and may result in errors that are difficult to estimate. The presented method is based on the simultaneous collection of both the image of backscattered electrons and elemental maps obtained using an energy dispersive X-ray analyzer. Multivariate comparative analysis of the matrix was used to characterize a technique of distinguishing epoxy resin from a hydrated cement-based substrate. The proposed algorithm significantly facilitates the detection of individual interface components, and also makes comparative analyzes repeatable and reliable.

## 1. Introduction

Scanning electron microscopy (SEM) with the use of backscattered electron contrast (BSE) is one of the leading methods of investigating and imaging multiphase, complex, cement-based structures. It has been used since the 1980s, when quantitative characterization of a microstructure was first used [1,2]. The popularization of SEM has led to the development of a fundamental method of porous segmentation (using the overflow method [3]), which derives from the fact that most non-scattered electrons correspond to voids in a material. The method is widely used today for the analysis of: lime mortars [4], the influence of additives on cement paste [5–8], the interface between fresh cement paste and recycled concrete aggregates [9], and the interface between the substrate and a coating [10]. Importantly, however, work [3] only dealt with the microstructure of cement paste, with the aggregate remaining in the concrete being masked manually. It is an approach that has become the so-called standard in civil engineering, both during the analysis of cement materials [5,11], and also during the analysis of resin coatings on a cement paste base [12]. There are similar problems with the X-ray computed tomography (microCT) technique, and therefore, methods of detecting individual components of the microstructure are intensively developed. These methods, however, require some operator intervention when processing data [13]. A more detailed assessment of the microstructure not only allows for the segmentation of porosity, but also for the recognition of the presence of individual microstructure components in SEM-BSE images based on different shades of grey [14]. Among the alternative methods of SEM image analysis, there are ones based on machine learning and SEM-BSE images [15], which also allow for an extended microstructural description. However, it should be remembered that no method offers a comprehensive description of a material's porosity. Porosimetries are usually limited to the recognition of open pores. In turn, microCT methods do not recognize pores on the nanometric scale, with SEM-BSE analyzes being limited by the resolution of the used device and the size of the taken micrographs. Pores smaller than one pixel of the analyzed image are able to be detected, and pores with dimensions exceeding several to several percent of the frame can be misinterpreted. The SEM-BSE method is therefore not suitable for an absolute quantitative description of a material, but is very helpful for comparative analyzes of a series of experimental and reference samples that are prepared and imaged in the same reproducible way.

The problem of manual masking of certain zones of images for the correct segmentation of porosity is especially important in the case of materials and samples with complex microstructures. In the following analysis, a sample of an epoxy floor on a concrete base was used as an extremely unfavorable example. Epoxy resin is characterized by excellent chemical resistance, good mechanical properties, and the fact that surfaces covered with it are easy to clean, which is why it is popular in industrial civil engineering. Unfortunately, however, epoxy resin is relatively expensive when compared to other floor coatings. It should be mentioned that the resin consists of very harmful substances, e.g. carcinogenic bisphenol A. It is therefore a common practice to add aggregates as fillers to the epoxy coating, which reduces the mass of the epoxy resin needed to make the coating, in turn lowering its cost and the content of harmful ingredients. At the same time, adding aggregates in the right amount (usually no more than 40% of the coating's mass) does not deteriorate the properties of the coating. Following the principle of sustainable development and the reuse of waste, research is currently being carried out on the use of recycled fine aggregates (RFA) as fillers for epoxy resin [16], especially as waste mineral aggregates constitute a significant part of all waste in Europe [17]. Therefore, in the authors' research, it was decided to add recycling aggregate to the epoxy resin. It should be added that such a choice of material was due to the fact that the issue of reducing the amount of epoxy filler in floors is an important and current problem. Moreover, the microstructure (within the multiphase between the substrate and the overlay) of a sample of this type is extremely difficult to analyze.

It is worth noting that the manual masking of the fragments of a microstructure in the case of the segmentation of porosity in interfacial transition zones (ITZ) is also an important problem. As a result of the bonding process of epoxy resins, it may turn out that porosities accumulate around RFAs or other reinforcements. The detection of this type of phenomenon allows for the improvement of the technology of the analyzed systems, as well as for further work aimed at creating more sustainable epoxy composites. The manual removal of individual areas during analysis may cause unmasked fragments of aggregate to be an unintentionally considered, or a too extensive part of an image (together with the surrounding section of porosity) to be removed.

The work focuses on the issue of distinguishing the individual components of the concrete-resin and concrete-concrete composite interface for the needs of a quantitative and qualitative description of its microstructure, especially with regards to local porosity within the joint. Our previous attempts [5,10] required the masking of the remaining elements of the structure. However, in the approach described below it was decided to extend the analysis to include the use of concentration maps of elements, which were obtained by X-ray energy dispersion spectroscopy (EDS) methods, and which were previously occasionally used to characterize the microstructure [7]. A method of comparing individual pairs of images was applied in order to find correlations between different chemical compositions, whereas the SEM-BSE image was used for a particularly precise delineation of components. An image analysis method and exemplary results obtained on various types of concrete and resin interfaces were presented. The proposed method is the first that allows the individual interface components on a micro and submicrometric scale to be distinguished, and also enables the precise analysis of porosity in the area of the connection of two materials with a complex morphology. It is also worth noting that methods based on gas and mercury porosimetry give general macroscopic information, and also that methods based on X-ray tomography cannot distinguish the local chemical composition of a sample or assign the segmented porosity to the corresponding structure component.

## 2. Materials and methods

The examined samples came from the epoxy resin coating that was previously described in [18]. The sample substrate consists of ready-mixed concrete composed of limestone powder, type I Portland cement, and quartz aggregate with a grain size of 0-4 mm. The water-to-cement ratio (w/c) was equal to 0.48. After 28 days of curing, the surface was ground, and a two-component epoxy resin composite (StoPox BB OS, Sto Ltd., Wroclaw, Poland) and RFA were then applied. The methodology of making the samples was close to that used for the commercial formation of epoxy floors on a concrete base. Recycled aggregate was added to the epoxy resin and mixed for about 3 minutes. A curing agent was then added to the mixture and mixed until a uniform consistency and color were obtained. The ready mix was spread on the surface of the substrate with a metal notched float for a time not exceeding 20 minutes from the moment of adding the curing agent. The weight ratio of the resin base, hardener, and RFA was 4:1:3. The epoxy-cement composite was cured for 7 days in a controlled laboratory environment at the temperature of 21°C and with a relative humidity of less than 60%. Extensive research on the influence of the participation of RFA on the properties of the obtained coating was described in [18]. Difficulties with the microstructural description of such a complex material were one of the motivations for the undertaken research. Epoxy resin, as the lightest component of the composite, is the darkest during SEM-BSE imaging, and the RFA and the substrate also differ in contrast, which in turn makes it difficult to easily assign an image fragment to a given microstructure component group. The analyzed microstructures of the sample are summarized in Figure 1. Figure 1a shows the coating area, with a significant complexity of the its microstructure. In the middle of the frame, there is the RFA, which consists not only of the aggregate, but also of the accompanying cement paste. Standalone aggregates not surrounded by cement paste can also be seen in the resin. All the images (along with additional data) from Figures 1a, b, and c were used for the tests. Image b focuses on the recognition of fine aggregate fragments in the RFA and the resin. Figure 1c shows the most interesting place - the proper interface between the concrete substrate and the RFA reinforced resin layer. Here the aggregates appear both in the substrate, as a standalone coating reinforcement, and in the RFA fragments surrounded by cement paste. Sub images differ in terms of brightness and contrast, which is typical for SEM imaging.

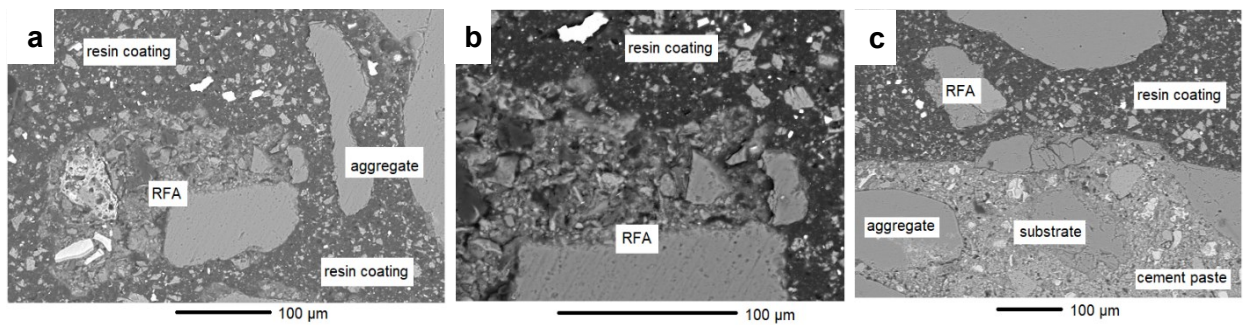

**Figure 1.** Microstructures of the different places of epoxy/RFA coating on the concrete substrate, which for further analysis were designated as: (a) sample 1, (b) sample 2, and (c) sample 3. Scanning electron microscopy, BSE detector, material contrast.

1 cm$^3$ cubes were cut from the samples, and they were then cold-mounted in epoxy resin, ground, and polished. The flat surface of a sample was connected to conducting copper tape and sputtered with a 40 nm layer of carbon in a high-vacuum evaporator. SEM imaging was performed using a JEOL JSM-6610A scanning electron microscope (Tokyo, Japan) using a two-field BSE detector, an accelerating voltage of 20 kV, a gun current of 40 nA, and a working distance of 10 mm. EDS analysis was performed using a JEOL JED-2200 detector that was installed on the used SEM. The objective aperture and beam parameters were selected in order to provide a count rate at a level of 7500 cps and a dead time not exceeding 10%. For each test, the data acquisition time was about 3 h, and elemental mappings were obtained. In these maps, the grey level linearly corresponds to the amount of a given element in each of the pixels of the image.

## 3 Experimental results

An image of backscattered electrons is first taken using material contrast in a scanning electron microscope. This image can be represented by an array $M = [m_{ij}] \in \mathcal{M}_{I \times J}$, where $m_{ij} \in [0,255]$ are pixels, and I and J are image dimensions in pixels. In turn, maps of the intensity of the occurrence of individual elements can be represented by matrices $E_X = [e_{ij}] \in \mathcal{M}_{I \times J}$, where X is the element symbol, and $e_{ij} \in [0,255]$ are pixels of the element's map. Notation $E_X(ij)$ denotes the element $ij$ of the matrix $E_X$.

To match the tones of the map to the corresponding parts of the sample, matrices $T_Q = [t_{ij}] \in \mathcal{M}_{I \times J}$ according to the formula $T_Q = \mathbb{1}_{\{(i,j): e_{ij} \in Q\}}$ were created for each map. Figure 2 illustrates the matrices for $Q := [a, b), a = 25k, b = 25(k+1), k = 0,1,\ldots,10$ for the element maps and BSE image of sample 2. The elements with the largest percentages in the sample (Si, Ca, Al, O) have a larger tonal spread in the relative abundance image than the samples with small percentages. As a result of the small amount of signal for the less abundant elements, some $T_Q$ maps are completely black.

For sample 2, it can be observed that Si map $T_{[0,25)}$ represents mainly resin, $T_{[25,150)}$ cement paste, and $T_{[150,255)}$ aggregate (Figure 2). Therefore, it can be assumed that parameters $p_1$ and $p_2$ approximately separate the areas of resin, cement paste, and aggregate. Parameter $p_1$ separates the resin from the cement paste, and $p_2$ - the cement paste and the aggregate. In addition, when using this approach for the Si map, Ti map, and BSE image, parameter $p_3$ can be determined, which separates the very bright, sulfur- and titanium-rich aggregate components from the other elements of the microstructure. Following the imperative of using the least noisy input data, the authors decided to indicate parameter p3 in the BSE image. The last parameter of the model was the parameter separating the calcium-rich aggregate from the resin. It can be found in the calcium map, and is denoted by $p_4$. The tonality of the maps of the elements was created in such a way that all the tonal availability of 8-bit files was used. For this reason, some maps have gaps in the histogram. This is because a maximum of $a$ counts were obtained in their case, and only $a$ shades of grey were available - distributed evenly in the 8-bit space. This phenomenon is visible in the Boolean maps of Fe, C, O, K, Ti, and S (Figure 2). This is an additional reason why it was decided to focus on the Ca and Si maps, which provide the most occupied shades of grey.

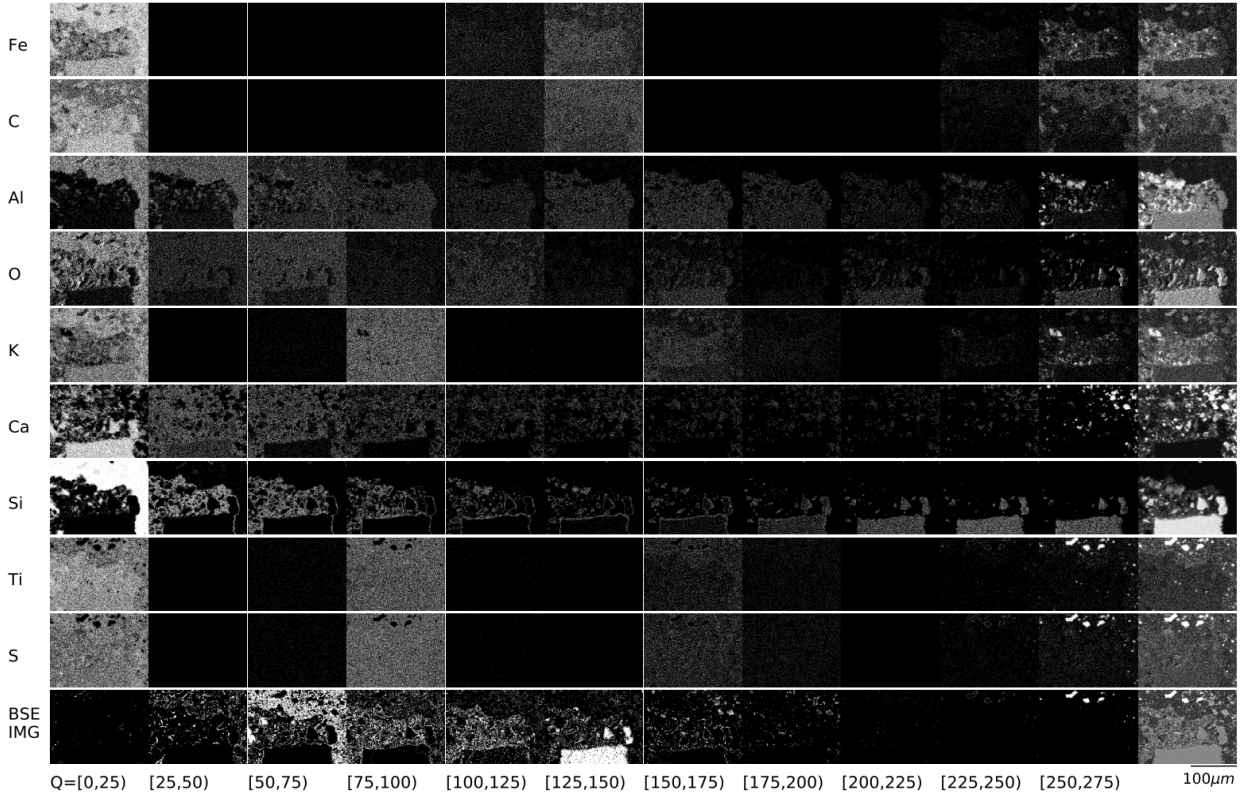

**Figure 2.** Boolean maps prepared for the element maps and BSE image of sample 2. The $Q$ range is the greyscale range of the pixels in the original image (shown on the far right).

Finally, the binary maps that designate the resin ($B_{resin}$, equation 1) and cement paste ($B_{cement}$, equation 2) can be determined from the image. The image of the resin separated from the other elements of the microstructure can be determined as $M \circ B_{resin}$, and the cement paste as $M \circ B_{cement}$, where $\circ$ is the Hadamard product symbol.

$$B_{resin} = \mathbb{1}_{Q_B},\ Q_B = \{(i,j): E_{Si}(ij) \leq p_1 \wedge M(ij) \leq p_3 \wedge E_{Ca}(ij) \leq p_4\} \tag{1}$$

$$B_{cement} = \mathbb{1}_{Q_C},\ Q_C = \{(i,j): p_i < E_{Si(ij)} \leq p_2 \wedge M(ij) \leq p_3\} \tag{2}$$

In order to find a method to reproducibly and objectively select the required parameters, the frequency histograms of the Si and Ca maps, as well as the BSE image, were analyzed.

The determination of parameter $p_1$ was performed differently - depending on what type of input data (file format) was dealt with. In the case of a file with the JPEG extension, the first derivative of the histogram was calculated, the first and the last value were changed to zero, and then, taking into account the arguments greater than the minimum argument, the first argument, for which the histogram value was greater than zero, was chosen. For a BMP extension, only non-zero values of the function were considered, the first derivative of the histogram was computed, the first and last values were changed to zero, the derivative was computed again, and the maximum of the function was then selected. The differences in the algorithms were due to the lossy compression of the sample image, which not only resulted in a change in the shape of the histogram, but also in the parameter being looked for to be shifted to a slightly different location. For this reason, it is recommended to work with lossless bitmap formats (BMP, TIFF) in the maximum available tonal range. Argument $p_1$ can therefore be considered as the place of the function's inflection, and this place can be determined analytically from the function's derivative.

The arguments of the first and last local maximum must be found, followed by the determination of a minimum between them. $p_2$ is the argument of this minimum (in the case of BMP extension, only non-zero values of the histogram are considered for the reasons described earlier). Parameter $p_3$ is the minimum argument among arguments larger than the argument of the first local maximum of the BSE image histogram, while parameter $p_4$ is determined like parameter $p_2$ using the Ca histogram instead of Si. The histograms of Ca and Si, the BSE images, and the determined parameters can be seen in Figure 3. Despite some differences in the shape of the plots, there are some common elements. The BSE histogram shows two typical peaks of resin (left for darker shades of grey) and aggregates (right from the middle of the histogram). However, there is another signal on both sides of the right peak, which is best seen at the lowest magnification of sample 3 (Figure 3c). This broad peak corresponds to differentiated cement paste phases, which can be distinguished from aggregates with the use of the elemental maps. The $p_3$ parameter allows the brightest areas of the image to be separated, and then allocated to aggregates, and not to cement paste. The Si histograms are divided almost in half by the $p_2$ parameter, which separates the local maximum on the right (typical for aggregates) from cement that contains less Si. In the case of further left, the $p_1$ parameter separates the Si-free resin from the rest of the structural components. The Ca histogram is dominated by Ca-free sites, which is typical for aggregates. The marked $p_4$ parameter separates the Ca-rich regions that stand out from the resin's structure (Figure 2).

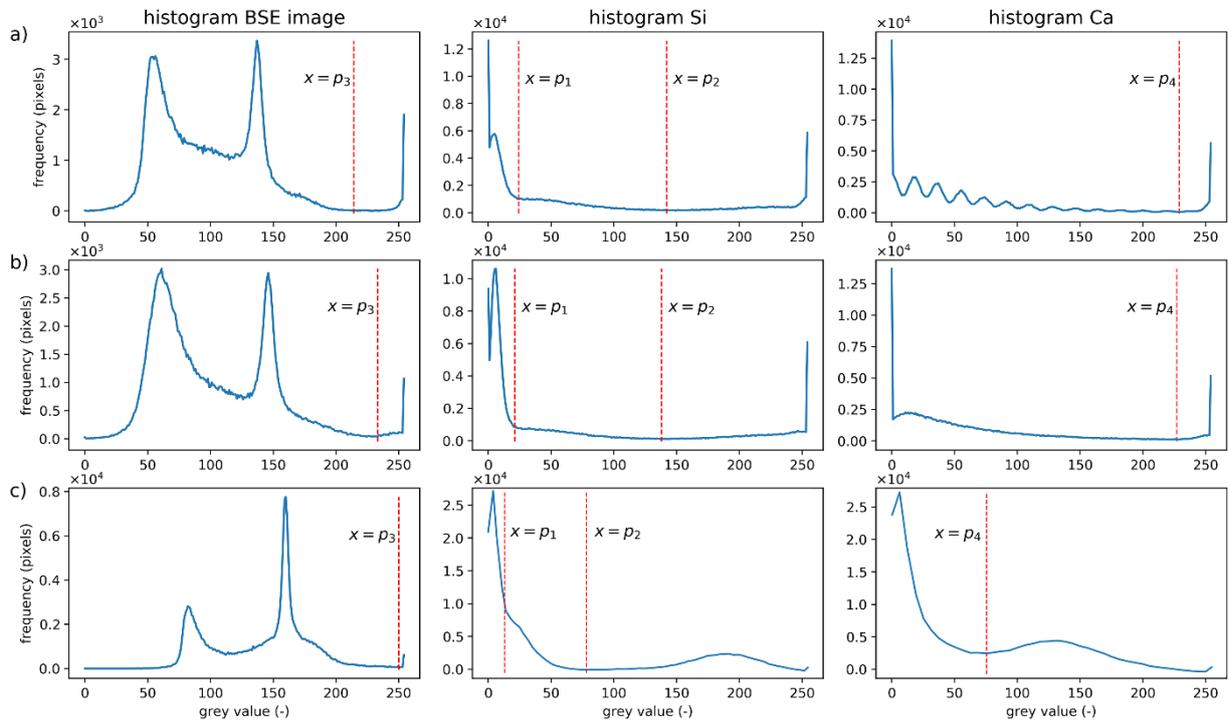

**Figure 3.** Histograms with parameters marked with red lines: (a) sample 1; (b) sample 2; (c) sample 3.

Anything that did not fall previously into the resin and cement paste categories was marked as aggregate. The analysis classified the structural components as expected. However, the results were significantly noisy, which was not intuitive. Logic dictates that there should not be many individual pixels of cement paste in the resin, and vice versa. Assuming that individual fractions are relatively continuous, the authors chose to denoise the results. The convolution of $B_{resin}$, $K(3 \times 3)$, and $K_{ij} = \frac{1}{9}$ was thresholded at 0.5 by formula $F_{resin} = M \circ \mathbb{1}_{\{(i,j):B_{resin}*K(ij)>0.5\}}$ in order to obtain a filtered image of the resin. The same operation was performed for matrix $B_{cement}$. The results of the division with filtration are shown in Figure 4. As can be seen, they were deprived of individual indications that are related to the noise of input signals. It is worth noting that the EDS signal is generated from a different depth than the BSE signal, and it is also noisy. Filtering allows for the reduction of noise at the expense of resolution, and the resolution is still limited by the physics of the formation of a signal in SEM.

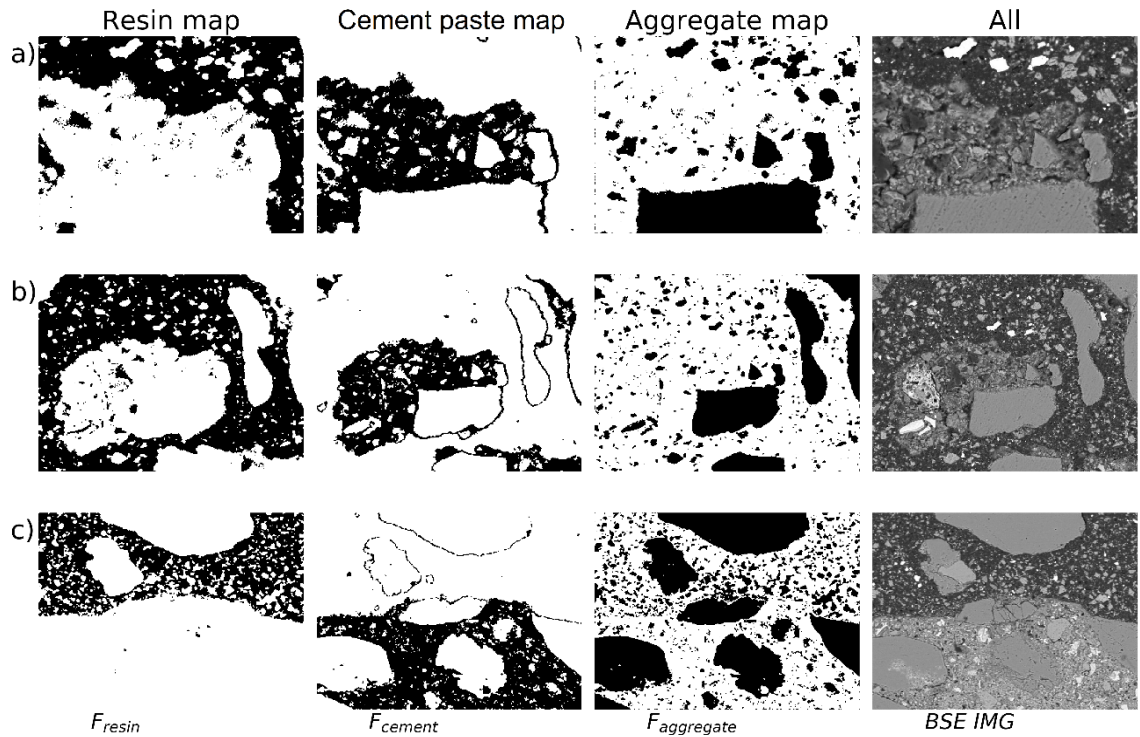

**Figure 4.** Division with filtration: (a) sample 1; (b) sample 2; (c) sample 3.

The used method allows for the recognition of resin, cement paste, and areas of aggregate. It is worth noting that the used Boolean map analysis method required an independent correlation of the occurrence of individual elements and shades in the SEM-BSE image, which then allowed for the selection of appropriate values for parameters $p_1$, $p_2$, $p_3$, and $p_4$ based on analytical mathematical functions. Therefore, the researcher does not have to use his own subjectivity in subsequent analyzes. It is worth emphasizing, however, that parameters $p_2$, $p_3$, and $p_4$ are located on relatively flattened fragments of the histograms (Figure 3), and therefore their shift by even 10 grey-value values does not visibly affect the obtained results. The only parameter for which a small shift would make the algorithm work much differently is $p_1$, which uses the Si content to separate resin and cement paste areas. Theoretically, the use of C instead of Si as the main resin component should be very promising, but the difficulty of its use is in the weak generation of the characteristic X-ray radiation by light elements. In the case of the Si map, there is a much larger signal-to-noise ratio, and to determine $p_1$, the function inflection method, which is used analogously to the determination of the porosity threshold for cement paste [3], is used. It is sensitive to changes in value, but can be analytically determined.

In addition, Figure 5 shows a plot of the generated porosity across the cross-section, with the porosity of the resin and the concrete substrate being distinguished. The plot was made using the modified Overflow method [3]. This method takes advantage of the fact that the threshold of the porosity can be determined at the first inflection of the cumulative histogram. It allows the porosity, regardless of the different brightness and contrast of different images, to be determined. In order to present the possibilities of the algorithm described above, the exemplary results of the analysis of the cross-sectional porosity of the place marked as sample 3 were shown. This is an image that normally could not be subjected to porosity analysis without manually masking the three components of the microstructure – cement paste, aggregate, and resin. Manual separation of structural components is subject to a significant bias in the subjectivism of the researcher, and due to the different brightness of cement paste and resin, the thresholds of both data sets fall in different places. By using a threshold appropriate for the resin, the porosity of the cement paste would be overestimated, and by using a threshold typical for cement paste, the measured porosity of the resin would be lower. In this case, the analysis matches each pixel with its corresponding structure component. For Figure 5b, the x-coordinate of the plot is the mean porosity in the given Y line (marked in Figure 5a). The orange line specifies the fraction of pores in pixels recognized as resin, the blue line represents the fraction of pores in pixels recognized as cement paste, and the dotted red plot represents cumulative porosity in areas not assigned to the aggregate (both resin and cement paste).

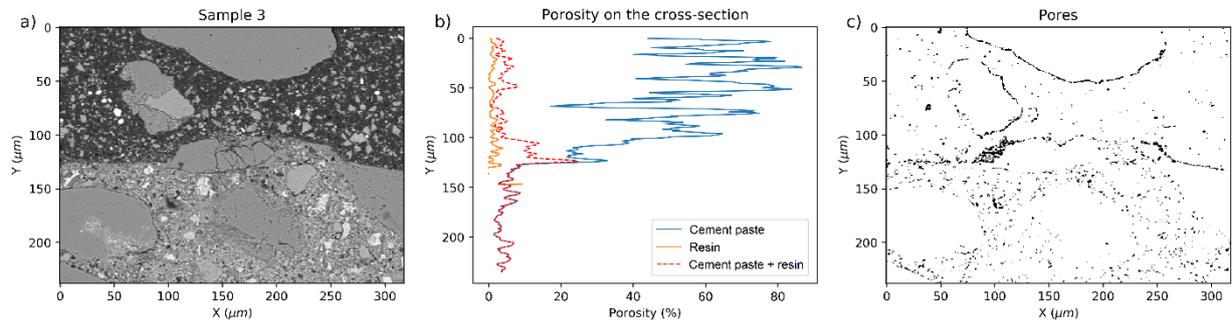

**Figure 5.** Porosity on the cross-section of sample 3: (a) BSE image; (b) porosity in the resin, porosity in the cement paste, and the total porosity of both materials; (c) segmented pores of the resin and cement paste.

The mean total porosity of the analyzed area is 5.08%, but the fraction of pores of the individual components of the structure is actually much more important. The sum of porosities in both materials is calculated by excluding the aggregate. The average porosity of the areas recognized as the resin is 1.11%, but at the boundary with the substrate (y value in Figure 5 between 100 and 130) it increases to 1.63%. The average fraction of the analyzed pores of the cement paste in the substrate is 8.55%, but is noticeably higher at the coating interface (y parameter in Figure 5 between 100 and 130), where it increases to 24.46%. In places where cement paste surrounds RFA fragments, its porosity often exceeds 50%, but this only applies to very narrow areas (Figure 5c). This is in line with typical values for porosity segmentation of epoxy coatings and cement-based materials [12].

By using the proposed algorithm, it was possible to separate resin porosity and cement paste porosity, which is something the standard method does not allow when the samples have different BSE contrasts [3]. The cross-sectional porosity for cement paste is calculated for each row as the number of cement paste pore pixels in the row divided by the number of cement paste pixels in the row, as is also the case for resin. The exceptionally high porosity of the cement paste regions in the resin coating may appear to be incorrect (as it is between 30-80%), especially as the porosity of cement paste alone is ten times less. This is in line with the prediction that RFA is covered with a cracked and porous layer of cement paste that provides a mechanical bond with the resin matrix. However, due to its viscosity, the resin is not able to fill all pores, and therefore the areas that are completely chemically identified in this region as cement paste have significant porosity. The resin-infiltrated areas do not show cement paste properties in chemical measurements, which qualifies them as solid parts of resin in the coating. Such an analytical result is due to the small amount of cement paste in the upper part of the image (mainly filled by resin), and the specific location of the cement paste (mainly at the RFA boundaries). In this location, cement paste is highly porous, and is responsible for the RFA/resin interface. This information was usually removed by manual masking of the microstructure's elements, or was interpreted as a disturbance in the masking process. However, reliance not only on the BSE signal, but also on EDS mappings has made it possible to unequivocally attribute this porosity to the thin layer of cement paste surrounding the RFA. This unique set of information from the ITZ area would not be obtainable without the simultaneous acquisition and analysis of BSE and EDS signals.

## 4. Conclusions

The goal of the research, which was to identify and mark the individual components of the microstructure of a multi-component composite based on cement paste and resin, was achieved. To accomplish this goal, logical maps of the occurrence of the specific intensity of elements were used, thanks to which it was possible to determine the model. Then, based on histograms of selected elements, the algorithm of selection of necessary parameters was specified. The used method made it possible to recognize local porosity, regardless of the brightness and contrast of the input image. The experimental information obtained on the ITZ of the concrete/resin+RFA composite allowed for the determination that the microporosities in the tested system accumulate in the immediate area of the substrate and coating connection, as well as on a thin cement paste layer surrounding the RFA. Further application of the method will include, inter alia, modification of the coatings in order to minimize the porosity around the RFA, and also the optimisation of the local microporosity and mechanical properties of the tested coatings.

Further development of the newly described method may involve the use of maps of other elements, especially Al or C (extending the use of elemental maps beyond the Ca that was already used), or the use of intensity quotients of two or more elements. In addition, other filtering methods can also be considered, including the disregarding of observations that occur too rarely in a given place. If the method is used for a different group of materials, the performing of Boolean maps for the SEM-BSE image and all detected chemical elements is suggested. This operation needs to be followed by recognition of the images and values in which the element of microcirculation that is important appears. In the last step, histograms for SEM-BSE, and their relevant maps, should be generated in order for them to be used to analytically determine the appropriate $p_n$ parameters that differentiate the microstructure elements. It is hoped that the proposed method will bring many benefits with regards to the analysis of not only cement-based materials, but also any multi-component composite coatings, as well as multiphase materials.

## Acknowledgments

The authors received funding from the project supported by the National Science Centre (NCN), Poland (grant no. 2020/37/N/ST8/03601), "Experimental evaluation of the properties of epoxy resin coatings modified with waste mineral powders (ANSWER)."

## Author statement

**Andrzej M. Żak**: Conceptualization, Formal analysis, Resources, Writing - Original Draft, Writing - Review & Editing, Supervision, Project administration.
**Anna Wieczorek**: Methodology, Software, Formal analysis, Investigation, Data Curation, Writing - Original Draft, Visualization.
**Agnieszka Chowaniec**: Validation, Writing - Review & Editing, Funding acquisition.
**Łukasz Sadowski**: Validation, Resources, Writing - Review & Editing, Supervision.